\documentclass{Interspeech2024}
\usepackage{multirow}
\usepackage{tikz}

\interspeechcameraready

\title{Can Synthetic Audio From Generative Foundation Models Assist Audio Recognition and Speech Modeling?}

\name[affiliation={1,}]{Tiantian}{Feng}
\name[affiliation={2}]{Dimitrios}{Dimitriadis}
\name[affiliation={1}]{Shrikanth S.}{Narayanan}

\address{
  $^1$University of Southern California, USA\\
  $^2$Amazon, USA}
\email{tiantiaf@usc.edu, dbdim@amazon.com, shrikann@usc.edu\thanks{This work was supported by USC-Amazon Center.}}


\begin{document}

\maketitle

\begin{abstract}
Recent advances in foundation models have enabled audio-generative models that produce high-fidelity sounds associated with music, events, and human actions. Despite the success achieved in modern audio-generative models, the conventional approach to assessing the quality of the audio generation relies heavily on distance metrics like Frechet Audio Distance. In contrast, we aim to evaluate the quality of audio generation by examining the effectiveness of using them as training data. Specifically, we conduct studies to explore the use of synthetic audio for audio recognition. Moreover, we investigate whether synthetic audio can serve as a resource for data augmentation in speech-related modeling. Our comprehensive experiments demonstrate the potential of using synthetic audio for audio recognition and speech-related modeling. Our code is available at \url{https://github.com/usc-sail/SynthAudio}.
\end{abstract}

\section{Introduction}
% Pre-trained speech models
The rapid developments in deep learning \cite{lecun2015deep} have advanced the landscape of AI and the reach of its applications, creating new frontiers in healthcare \cite{miotto2018deep}, virtual assistants \cite{chen2019deep}, and content generation \cite{bommasani2021opportunities}. More recently, we have witnessed waves of foundation models \cite{bommasani2021opportunities}, enabling enormous breakthroughs in generative models such as ChatGPT and Gemini. As generative models continue to evolve in natural language processing \cite{touvron2023llama} and computer vision \cite{croitoru2023diffusion}, researchers have also released various audio generative models showing their ability to synthesize realistic audio content. For example, AUDIOGEN \cite{kreuk2022audiogen} is a recently released auto-regressive audio generative model that can generate high-quality audio based on the text description of a sound. Likewise, AudioLDM \cite{liu2023audioldm} is a text-to-audio system that generates audio based on the diffusion technique \cite{ho2020denoising}.

% metrics
While these models hold promise in creating high-fidelity sounds, one fundamental challenge is to ensure generation quality, which includes accuracy, consistency, and diversity. The conventional approach to evaluating the audio generation quality is through distance metrics like Frechet Audio Distance (FAD) \cite{kilgour2019frechet}. This approach uses a pre-trained audio classification model to obtain the posteriors of synthetic and real audio, which is then used to calculate the KL divergence to quantify the generation quality. Although this evaluation aligns with the training objective of audio generative models, this metric fails to indicate how well the audio generation matches the real audio sound. Apart from FAD, studies also attempt to include humans in rating the relevance of audio generation based on a given text. However, this approach requires substantial human efforts.

% What we do
\vspace{0.4mm}
\noindent \textbf{Our Objectives:} Unlike previous works that evaluate audio generation quality through distance measures or human annotations, we argue that high-fidelity audio generations can also serve as the data source for training and augmentation. Inspired by prior work on synthetic images for image recognition \cite{he2022synthetic}, we investigate if synthetic audio can assist audio recognition and speech-related modeling. Specifically, we compare the training performance using synthetic audio generated from several popular audio generative models (AUDIOGEN \cite{kreuk2022audiogen}, AudioLDM2 \cite{liu2023audioldm}, and MusicGen \cite{copet2024simple}) involving general audio, music, and human action sounds. Moreover, we investigate whether synthetic audio is a feasible source of data augmentation to improve speech modeling performance, including speech emotion recognition and keyword spotting. Notably, SynthASR \cite{fazel2021synthasr} is a similar work to ours, but it focuses on using synthetic speech to improve ASR. 
Our findings are summarized below:

\begin{figure*}[t]
	\centering
	\includegraphics[width=0.98\linewidth]{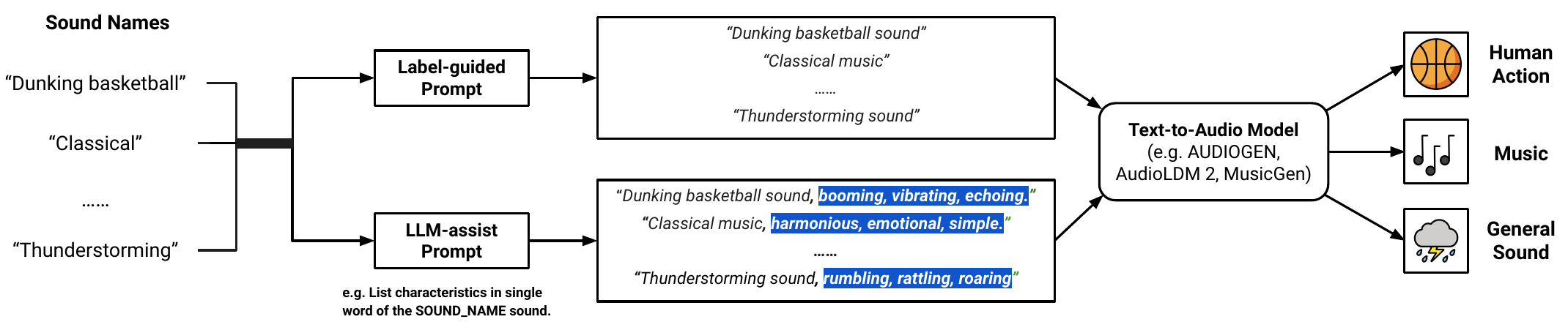}
    \vspace{-1.5mm}
    \caption{The audio generation approaches in this work: label-guided and LLM-assisted prompts. The label-guided prompt creates the prompt based on the label names, while the LLM-assisted prompt augments the label names with sound descriptions using LLMs. We use images from \url{https://openmoji.org/}}
    \vspace{-3.5mm}
    \label{fig:data_generation}
\end{figure*}

\begin{itemize}[leftmargin=*]

    \item \textbf{Zero-shot Audio Recognition:} Different audio generative models show unique qualities in their output, as measured by the effectiveness of using the generated audio for modeling. Here, the zero-shot (only training on synthetic data) is the same as in \cite{he2022synthetic}. We identify that MusicGen yields the best zero-shot music genre predictions, and AUDIOGEN delivers the best zero-shot accuracy on remaining audio predictions.
    
    \item \textbf{Factors Impacting Zero-shot Audio Recognition:} We analyzed factors related to audio generation that impact the zero-shot performance, showing that augmenting the prompt with sound descriptions from LLMs and increasing the generation quantity improves audio recognition.
    
    \item \textbf{Mixed Training For Audio Recognition:} We mix synthetic and real audio for model training. The audio recognition benefits significantly when combining synthetic audio in settings with limited real audio, but this does not hold when audio samples are adequate.
    Motivated by the data-centric approach \cite{motamedi2021data}, we filter out synthetic audio with lower performance in audio recognition, leading to increased performance when combining synthetic audio with complete real audio.
    
    \item \textbf{Data augmentation for Speech Modeling:} 
    We compare synthetic and real audio as a source of data augmentation in speech modeling and show that synthetic audio achieves competitive performance Moreover, speech models augmented with synthetic audio increase the robustness against environmental noises in speech-related classification tasks. 
    
\end{itemize}

\section{Audio Generative Models}
In this paper, we mainly investigate three popular audio generative models: AUDIOGEN, AudioLDM 2, and MusicGen. We particularly used MusicGen for music generation while we performed the generation of general sounds, human action sounds, and music with AUDIOGEN and AudioLDM 2.

\vspace{0.3mm}
\noindent \textbf{AUDIOGEN:} AUDIOGEN is an auto-regressive audio generative model that involves audio representation learning through the reconstruction objective. Following the learning of audio representation, AUDIOGEN trains the audio-language model to generate audio conditioned on text.

\vspace{0.3mm}
\noindent \textbf{AudioLDM 2:} AudioLDM 2 proposes learning the latent audio representation using the audio-masked autoencoder (AudioMAE \cite{huang2022masked}). The authors apply the GPT-2 \cite{radford2019language} to align different modalities with AudioMAE features. Finally, the Latent Diffusion Model is used to train the audio generation. 

\vspace{0.3mm}
\noindent \textbf{MusicGen:} MusicGen is a music generation model based on an auto-regressive transformer-based decoder \cite{vaswani2017attention}. Unlike conventional music generative models, MusicGen uses a simple but effective codebook interleaving and performs text and melody conditioning to generate high-quality music samples.

\section{Audio Generation and Recognition}

\subsection{Audio Dataset Generation}

We present the audio generation process in this work in Figure~\ref{fig:data_generation}. Our audio generation process is primarily inspired by prior works in \cite{he2022synthetic, shipard2023diversity, feng2024can, li2024synthetic, zhang2023gpt}. Here, we investigate two types of text prompts: label-guided and LLM-assisted prompts.

\vspace{0.3mm}
\noindent \textbf{Label-guided Prompt:} Label-guided prompt is a simple method that crafts the text prompt with only the label. For example, given a sound class of dunking basketball, we can create the prompt message as "\texttt{Dunking basketball sound}." This approach has been shown to provide promising zero-shot performance in image recognition by \cite{shipard2023diversity, zhang2023gpt}.
% Meanwhile, the prompt text associated with classical music is "\texttt{Classical music}.".

\vspace{0.3mm}
\noindent \textbf{LLM-assisted Prompt:} One challenge with label-guided prompts is limited diversity owing to constrained semantics. Several prior works have shown that increasing generation diversity \cite{shipard2023diversity, he2022synthetic} consistently improves zero-shot image recognition performance. To increase the diversity in audio generation, we leverage the LLM to provide characteristics of a given sound to augment the label-guided prompt. Here, we provide an example of an LLM-assisted prompt below:

\noindent \texttt{\textbf{User:} List three characteristics of classical music each in a single word.}

\noindent \texttt{\textbf{LLM:} rhythmic; balanced; harmonious.}

\noindent \texttt{\textbf{LLM-assisted prompt:} Classical music, rhythmic, balanced, harmonious.}

\subsection{Audio Recognition Modeling}

Once we obtain the synthetic audio dataset, we perform audio recognition using the pre-trained audio models. Specifically, we choose the Self-Supervised Audio Spectrogram Transformer (SSAST) \cite{gong2022ssast} as our audio modeling backbone. The SSAST is pre-trained to predict the masked spectrogram patch on audio samples, leading to competitive performances in multiple popular audio recognition testbeds.

\begin{table}[t]
    % \caption{Summary of dataset statistics used in this work. We select 20 and 25 soundable human actions in UCF101 and ActivityNet datasets, respectively.}
    \caption{Summary of dataset statistics used in this work. We select 20 soundable human actions in UCF101.}
    \vspace{-2mm}
    \centering
    \small
    \begin{tabular*}{0.9\linewidth}{lccc}
        \toprule
        
        \multirow{1}{*}{\textbf{Datasets}} & 
        \multirow{1}{*}{\textbf{Audio Domain}} & \multirow{1}{*}{\textbf{Classes}} & \multirow{1}{*}{\textbf{Data Size}}  \\
        \midrule

        \textbf{ESC50} & General Sound & 50 & 2,000 \\ 
        \midrule
        \textbf{GTZAN} & Music & 10 & 1,000 \\
        \midrule
        
        \textbf{UCF101} & \multirow{2}{*}{Human Action} & 20 & 2,840 \\ 
        
        \textbf{ActivityNet} & & 20 & 1,760 \\
    
        \bottomrule
    \end{tabular*}
    \vspace{-4mm}
    \label{table:datasets}
\end{table}

\section{Audio Recognition Experiments}

\subsection{Audio Recognition Datasets}

In this work, we choose ESC50 \cite{piczak2015esc} for general sound classification, GTZAN \cite{george2001automatic} for music genre classification, and UCF101 \cite{soomro2012ucf101} and ActivityNet \cite{caba2015activitynet} for human action recognition. A detailed overview of the datasets is provided in Table~\ref{table:datasets}. The ESC-50 dataset covers broad sound categories, including animal, nature, human, indoor, and urban sounds, for general sound classification. We chose this dataset over others like AudioSet \cite{gemmeke2017audio}, FSD50K \cite{fonseca2021fsd50k}, and VGG-Sound \cite{chen2020vggsound}, as it has not been used in training existing audio generative models. Moreover, GTZAN is used for music genre classification featuring ten music genres, each with 100 data samples. Lastly, UCF101 and ActivityNet are human action recognition datasets sourced from YouTube videos. Here, we manually curated subsets of 20 and 20 acoustically representative human actions from UCF101 and ActivityNet, respectively, as many action classes (e.g., apply makeup) are not directly recognizable from audio. The class names are presented in our GitHub repo\footnote{We thank Barrett Wang for assisting with the label selection}.

\subsection{Audio Generation}
We perform the audio generation on AUDIOGEN and AudioLDM 2 with all sound categories, while we only use MusicGen to generate music. We generated 30, 60, 150 audio data samples per general sound class, human action sound class, and music genre class, respectively. The generation number approximately matches the training data size in each dataset. In LLM-assist prompt generation, we prompt the LLM to generate ten characteristics given a sound class, and we randomly sample three characteristics in crafting prompts for audio generation. The LLM used in this work is Gemini-1.0. We generate the music in 10 seconds, and remaining sound classes in 5 seconds.

\subsection{Training Details}
We choose the pre-trained SSAST-Base model with a masked patch of size 400 for training audio recognition, as this model delivers the best audio recognition performance based on \cite{gong2022ssast}. All audio is resampled to a mono channel at 16K Hz. For baseline training on ESC50 and GTZAN datasets, we apply 5-fold cross-validation. We train UCF101 using cross-validation with its standard 3-fold splits while experimenting with ActivityNet following the standard train, validation, and test split. As the audio in GTZAN is 30 seconds long, we cropped the middle 10 seconds for the training music genre classification, and for other experiments, we limited the audio duration to 5 seconds. Each experiment adopts a learning rate from ${0.0001, 0.0005}$ and a maximum training epoch of 30, consistently applied to both synthetic and real audio training. Average accuracy is reported by averaging accuracy across different training folds.

\begin{table}[t]
    \caption{Comparing real data training with zero-shot synthetic audio training using different audio generative models based on the label-guided prompt, and the metric is accuracy.}
    \vspace{-2mm}
    \centering
    \footnotesize
    \begin{tabular*}{\linewidth}{p{1.1cm}cccc}
        \toprule
        
        \multirow{1}{*}{\textbf{Datasets}} & 
        \multirow{1}{*}{\textbf{Real}} & \multirow{1}{*}{\textbf{AudioLDM 2}} & 
        \multirow{1}{*}{\textbf{AUDIOGEN}} & \multirow{1}{*}{\textbf{MusicGen}}  \\
        \cmidrule(lr){1-1} \cmidrule(lr){2-2} \cmidrule(lr){3-5}

        \textbf{ESC50} & 88.6 & 26.7 & \textbf{48.3} & - \\ 
        \textbf{GTZAN} & 76.4 & 18.9 & 20.4 & \textbf{20.6} \\
        \textbf{UCF101} & 72.2 & 25.2 & \textbf{30.3} & - \\ 
        \textbf{ActivityNet} & 30.6 & 19.2 & \textbf{20.6} & - \\
    
        \bottomrule
    \end{tabular*}
    \vspace{-2.5mm}
    \label{table:zeroshot}
\end{table}

\section{Audio Recognition with Synthetic Audio}

\subsection{How Good/Bad is Label-guided Zero-shot Accuracy?}

We first explore the performance of the label-guided zero-shot audio recognition, as shown in Table~\ref{table:zeroshot}. Specifically, we compare training with real audio to training with synthetic audio using various audio generative models. The comparisons reveal a consistent and significant underperformance of zero-shot learning with synthetic audio compared to real audio training. AUDIOGEN consistently outperforms AudioLDM 2, while MusicGen performs best in zero-shot music genre classification. Due to the high variability in generation quality, we proceed with the remaining analysis using the audio generative models that show the best zero-shot performance for each dataset.

%  with the generation model based on the best performance from the label-guided prompt
\begin{table}[t]
    \caption{Comparing synthetic audio training between label-guided and LLM-assisted prompt. The metric is accuracy.}
    \vspace{-2mm}
    \centering
    \footnotesize
    \begin{tabular*}{0.75\linewidth}{p{1.3cm}cccc}
        \toprule
        
        \multirow{1}{*}{\textbf{Datasets}} & \multirow{1}{*}{\textbf{Label-Guided}} & 
        \multirow{1}{*}{\textbf{LLM-Assisted}}  \\
        \midrule

        \textbf{ESC50} & 48.3 & \textbf{57.0} \\ 
        \textbf{GTZAN} & 20.6 & \textbf{48.2} \\
        \textbf{UCF101} & 30.3 & \textbf{36.5} \\ 
        \textbf{ActivityNet} & 20.6 & \textbf{23.6} \\
    
        \bottomrule
    \end{tabular*}
    \vspace{-3mm}
    \label{table:zeroshot_llm}
\end{table}

\subsection{Can LLMs Improve Zero-shot Audio Recognition?}

We further investigated the zero-shot audio recognition through the LLM-assisted prompt. We adopt MusicGen in music generation and AUDIOGEN for the remaining audio generation. We present the performances of two zero-shot audio recognition training in Table~\ref{table:zeroshot_llm}. The comparisons demonstrate that zero-shot audio recognition using synthetic audio generated through the LLM-assisted prompt yields higher performance than the label-guided prompt. This indicates the importance of enriching semantic diversity in crafting text prompts for audio generation.

\begin{figure}[t]
	\centering
	\includegraphics[width=\linewidth]{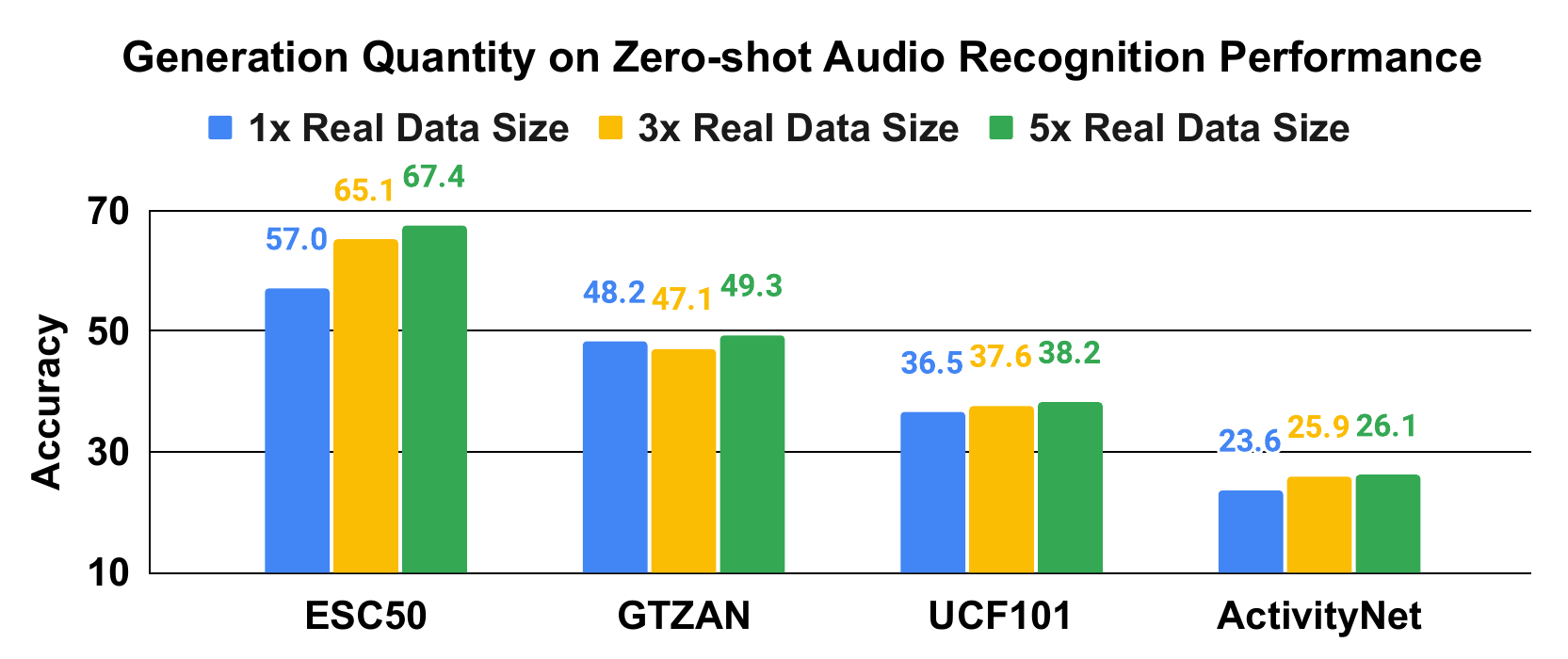}
    \vspace{-6mm}
    \caption{Comparisons of zero-shot audio recognition varying generation quantities. We study the generation quantity of approximately 1x, 3x, and 5x of the original training audio size.}
    \vspace{-3mm}
    \label{fig:quantity}
\end{figure}

\begin{figure}[t] {
    \centering
    
    \begin{tikzpicture}

        \node[draw=none,fill=none] at (0,5.){\includegraphics[width=\linewidth]{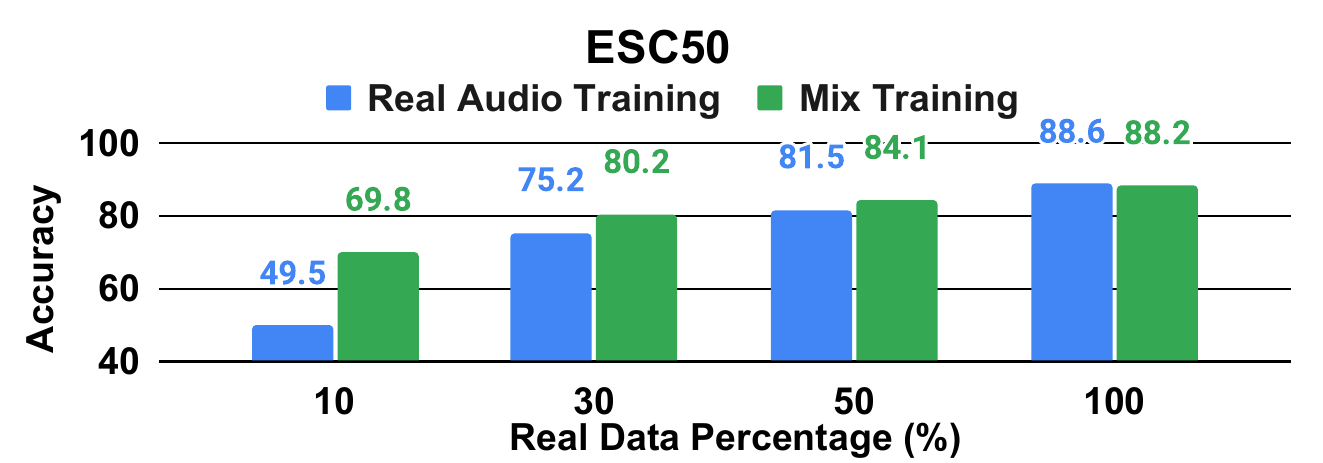}};

        \node[draw=none,fill=none] at (0,2.5){\includegraphics[width=\linewidth]{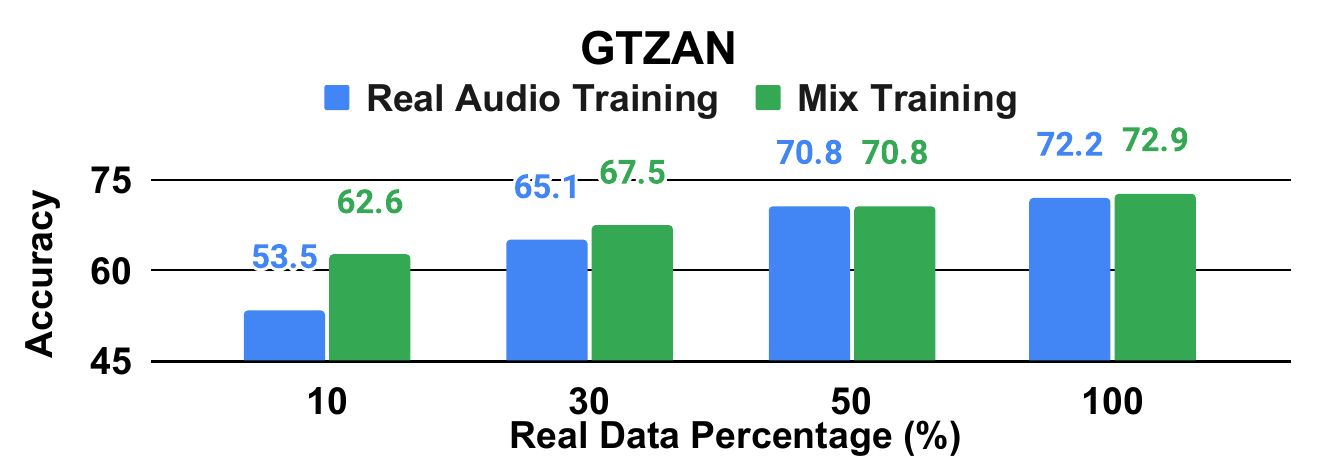}};
        
        \node[draw=none,fill=none] at (0,0){\includegraphics[width=\linewidth]{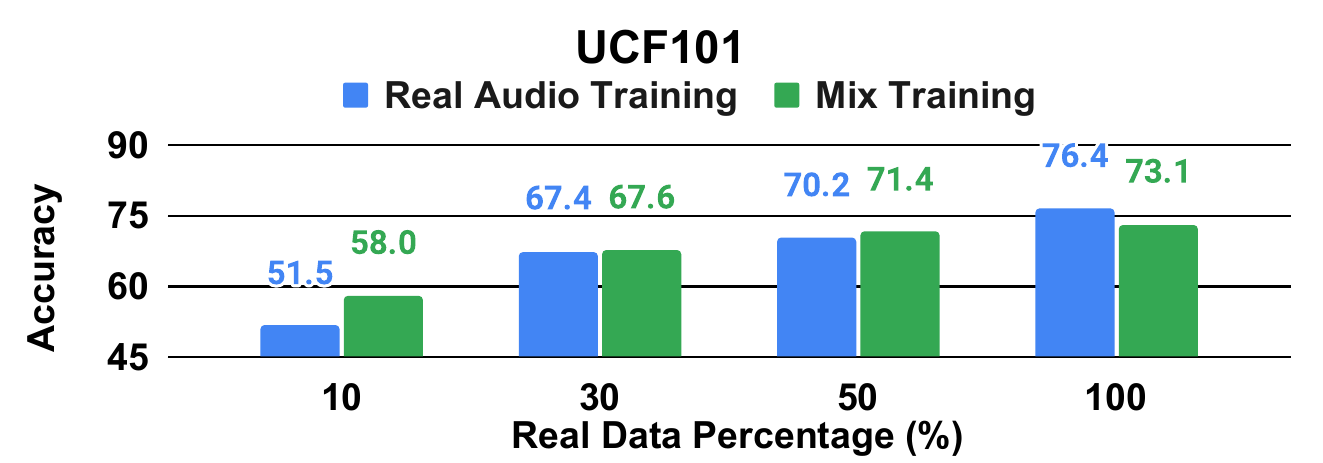}};

    \end{tikzpicture}
    \vspace{-7mm}
    \caption{Performance comparisons between real audio training and mixed training at different read data percentages.}
    \label{fig:mix_training}
    \vspace{-3.5mm}
} \end{figure}

\begin{figure}[t]
	\centering
	\includegraphics[width=\linewidth]{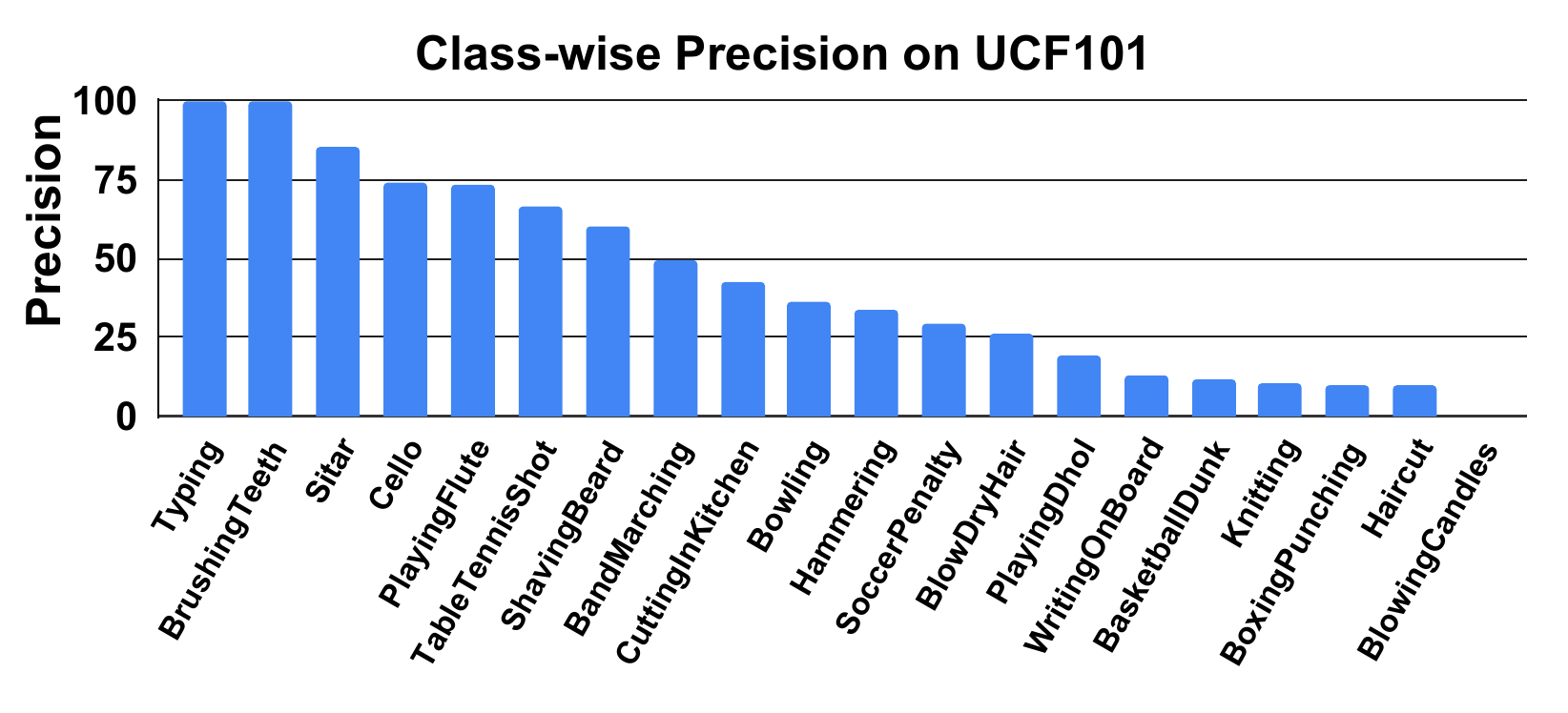}
    \vspace{-7mm}
    \caption{Class-wise precision of the zero-shot audio recognition on UCF101 data.}
    \vspace{-3.5mm}
    \label{fig:classwise_ucf101}
\end{figure}

\subsection{Can Generation Quantity Impact Zero-shot Accuracy?}

In addition to prompt diversity, we explore how varying the quantity of audio generation impacts zero-shot audio recognition performance. We generate sounds at different numbers per sound class—$\{30, 90, 150\}$ for general sound, $\{60, 180, 300\}$ for human action sound, and $\{100, 300, 500\}$ for music, approximating 1x, 3x, and 5x the real training audio size. Results in Figure~\ref{fig:quantity} show that increased generation consistently improves zero-shot audio recognition. However, these improvements tend to saturate with generation size to reach 5x the real audio size.

\subsection{Can Synthetic Audio Assist Audio Recognition?}

We further investigate whether synthetic audio enhances audio recognition with real audio. To answer this, we perform the training experiments that combine the real audio with synthetic audio. Specifically, we study the mix training with real audio ratio in $\{10\%, 30\%, 50\%, 100\%\}$. Here, the real audio ratio of $100\%$ represents mixing training with complete real audio and synthetic audio. Here, we constrain the synthetic audio size to 1x the complete real audio size in all experiments. Figure~\ref{fig:mix_training} shows that mixed training substantially improves recognition, particularly with limited real audio. For example, we observe a $20\%$ accuracy increase in the ESC50 dataset when there is only $10\%$ real audio. Performance gains are consistent across datasets with real data percentages of $\{10\%, 30\%, 50\%\}$. However, mixed training with complete real audio underperforms complete real audio training in multiple datasets.

\begin{table}[t]
    \caption{Comparing real audio training with data-driven mixed training. Complete real audio is presented in both experiments.}
    \vspace{-2mm}
    \centering
    \footnotesize
    \begin{tabular*}{\linewidth}{p{1.3cm}cccc}
        \toprule
        
        \multirow{1}{*}{\textbf{Datasets}} & \multirow{1}{*}{\textbf{Real Audio Training}} & 
        \multirow{1}{*}{\textbf{Data-centric Mix Training}}  \\
        \midrule

        \textbf{ESC50} & 88.6 & \textbf{89.9} \\ 
        \textbf{GTZAN} & 76.4 & \textbf{77.6} \\
        \textbf{UCF101} & 72.2 & \textbf{74.0} \\ 
        \textbf{ActivityNet} & 30.6 & \textbf{36.7} \\
    
        \bottomrule
    \end{tabular*}
    \vspace{-3.5mm}
    \label{table:data_centric}
\end{table}

\subsection{How to Boost Mixed Training for Audio Recognition?}

Results in Figure~\ref{fig:mix_training} show that a simple combination of synthetic and real audio does not consistently improve audio recognition when sufficient amounts of real audio are available for training. To understand the challenges in mixing synthetic and real audio for training, we adopt a data-centric concept to analyze the classwise precision in zero-shot audio recognition using the UCF101 dataset as an example in Fig~\ref{fig:classwise_ucf101}. We identify that many classes struggle to achieve $25\%$ precision for audio recognition, indicating their lower generation precision. Therefore, during mixed training, we filter out classes with bottom $20\%$ zero-shot precision in each dataset. Table~\ref{table:data_centric} compares data-centric mixed training with complete audio training, showing consistent improvement across all datasets. This underscores the importance of auditing synthetic audio quality by filtering out challenging sound generations for effective mixed training.

\section{Speech Modeling with Synthetic Audio}

In this section, we shift focus from general audio recognition to speech-related modeling, exploring the potential of synthetic audio as a data augmentation source for speech modeling. Specifically, we perform speech modeling in speech emotion recognition (IEMOCAP \cite{busso2008iemocap}) and keyword spotting (GCommands \cite{warden2018speech}). We use Whisper-Base \cite{radford2023robust} and SSAST-Tiny \cite{gong2022ssast} as backbones for training IEMOCAP and GCommands, respectively. We follow the guidelines in \cite{feng2023foundation} and \cite{gong2022ssast} in training IEMOCAP and GCommands, respectively. For the augmentation, we mix the speech with real or synthetic audio with SNR in 3-30 dB, involving sound classes from ESC50, and results are reported in unweighted average recall (UAR) and F1 scores for IEMOCAP and GCommands, respectively.

\begin{table}[t]
    \caption{Comparing speech modeling without augmentation (No Aug), with real audio augmentation (Real-Aud Aug), and with synthetic audio augmentation (Syn-Aud Aug). }
    \vspace{-2mm}
    \centering
    \footnotesize
    \begin{tabular*}{\linewidth}{p{1.2cm}p{0.65cm}ccc}
        \toprule
        
        \multirow{1}{*}{\textbf{Datasets}} & \multirow{1}{*}{\textbf{Metric}} & \multirow{1}{*}{\textbf{No Aug}} & 
        \multirow{1}{*}{\textbf{Real-Aud Aug}} & \multirow{1}{*}{\textbf{Syn-Aud Aug}} \\
        \midrule

        \textbf{IEMOCAP} & \multicolumn{1}{c}{Acc} & 68.6 & 68.7 & \textbf{69.1} \\ 
        \textbf{Gcommands} & \multicolumn{1}{c}{F1} & 0.959 & \textbf{0.964} & 0.960 \\

        \bottomrule
    \end{tabular*}
    \vspace{-3.5mm}
    \label{table:speech_modeling}
\end{table}

\begin{figure}[t] {
    \centering
    
    \begin{tikzpicture}

        \node[draw=none,fill=none] at (0,2.4){\includegraphics[width=\linewidth]{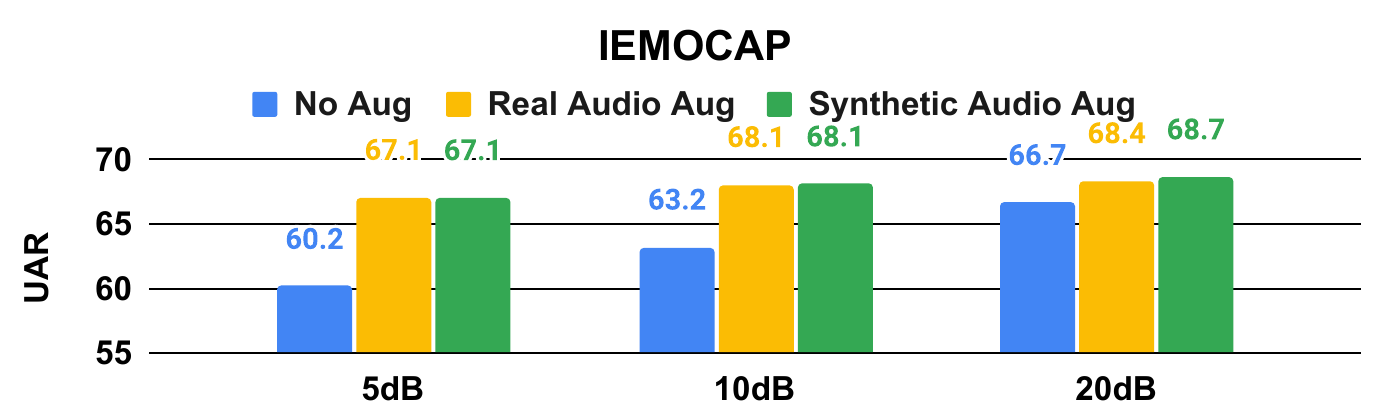}};

        \node[draw=none,fill=none] at (0,0){\includegraphics[width=\linewidth]{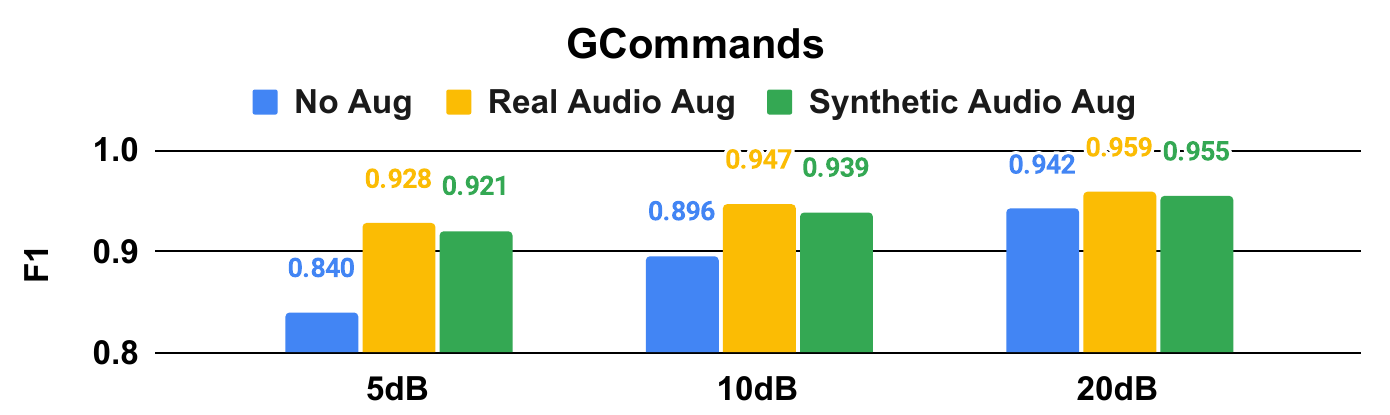}};

    \end{tikzpicture}
    \vspace{-6.5mm}
    \caption{Performances of speech training against real audio noises at 5, 10, and 20dB with different augmentation sources.}
    \label{fig:noisy_robust}
    \vspace{-3.5mm}
} \end{figure}

\subsection{Can Synthetic Audio Augment Speech Modeling?}

In Table~\ref{table:speech_modeling}, we compare speech modeling without augmentation, with real audio augmentation, and with synthetic audio augmentation. Results show that synthetic audio augmentation consistently improves speech modeling compared to no augmentation. Moreover, synthetic augmentation demonstrates competitive performance with real audio augmentation, even outperforming real audio augmentation on the IEMOCAP dataset.

\subsection{Can Synthetic Audio Improve Noise Robustness?}

Augmenting speech with audio in speech modeling is frequently used to increase model robustness against environmental noises. Therefore, we explore whether synthetic audio augmentation increases the noisy robustness of speech modeling against real audio noises. To answer this, we perform the inference on the test samples that are mixed with real audio noises. Specifically, we mix the test speech with real audio samples from ESC50 in the evaluation, and we test the model performance at the SNR of $\{5dB, 10dB, 20dB\}$. Results in Figure~\ref{fig:noisy_robust} show that synthetic audio offers an effective noise source to improve noise robustness in speech modeling. In particular, we observe that augmentation with synthetic audio yields similar performance to real audio even at SNR=5dB.

\section{Conclusion}
In this work, we study whether synthetic audio from generative foundation models can assist audio recognition and speech modeling. Our results show that synthetic audio presents comparable zero-shot performance using LLM-assisted prompts. Moreover, we show that mixed training with synthetic audio consistently benefits audio recognition when only limited real audio data is available. We further reveal that combining high-quality audio generation with real audio in a data-driven way consistently improves audio recognition. Finally, we show that synthetic audio is a useful source of data augmentation for speech-related modeling, improving both speech modeling performance and its noise robustness. In the future, we plan to extend our investigation of synthetic audio to multimodal learning involving audio modalities. Moreover, we plan to study the domain mismatch between synthetic audio and real audio.

%\ifinterspeechfinal
%     The Interspeech 2024 organisers
% \else
%     The authors
% \fi

\bibliographystyle{IEEEtran}
\bibliography{mybib}

\end{document}